\documentclass[preprintnumbers,superscriptaddress,showpacs,pra]{revtex4}
\usepackage{amsmath}
\usepackage{epsfig}
\usepackage{graphicx}
\usepackage{color}

\input{tcilatex}
\begin{document}

\title[Geometric Potential and Dirac Quantization]{Geometric Potential
Resulting from Dirac Quantization}
\author{D. K. Lian, L. D. Hu}
\affiliation{School for Theoretical Physics, School of Physics and Electronics, Hunan
University, Changsha 410082, China}
\author{Q. H. Liu}
\email{quanhuiliu@gmail.com}
\affiliation{School for Theoretical Physics, School of Physics and Electronics, Hunan
University, Changsha 410082, China}
\affiliation{Synergetic Innovation Center for Quantum Effects and Applications (SICQEA),
Hunan Normal University,Changsha 410081, China}
\date{\today }

\begin{abstract}
A fundamental problem regarding the Dirac quantization of a free particle on
an $N-1$ curved hypersurface embedded in $N$($\geq 2$) flat space is the
impossibility to give the same form of the curvature-induced quantum
potential, the geometric potential as commonly called, as that given by the
Schr\"{o}dinger equation method where the particle moves in a region
confined by a thin-layer sandwiching the surface. We resolve this problem by
means of previously proposed scheme that hypothesizes a simultaneous
quantization of positions, momenta, and Hamiltonian, among which the
operator-ordering-free section is identified and is then found sufficient to
lead to the expected form of geometric potential.
\end{abstract}

\maketitle

\section{Introduction}

For a free particle constrained to live on a curved surface or a curved
space, there is a curvature-induced potential in general, which however was
born with some problems concerning its forms. DeWitt in 1957 used a specific
generalization of Feynman's time-sliced formula in Cartesian coordinates and
found a surprising result that his amplitude turned out to satisfy a Schr%
\"{o}dinger equation different from what had previously assumed by Schr\"{o}%
dinger \cite{Sch} and Podolsky \cite{podolsky}. In addition to the kinetic
energy which is Laplace-Beltrami operator divided by two times of mass, his
Hamiltonian operator contained an extra effective potential proportional to
the intrinsic curvature scalar. For a particle constrained on ($N-1$%
)-dimensional smooth curved surface $\Sigma ^{N-1}$ in flat space $R^{N}$ ($%
N\succeq 2$), Jensen and Koppe\ in 1972 \cite{jk} and subsequently da Costa 
\cite{dacosta} in 1981 and 1982 developed the confining potential formalism
(CPF) (also known as the thin-layer quantization) to deal with the free
motion on the curved surface and demonstrated that the particle experiences
a quantum potential that depends on both the square of the trace of, and the
trace of the square of, the extrinsic curvature tensor of the curved
surface, which was later called the \textit{geometric potential }\cite{2004}%
. By the CPF we mean to write the Schr\"{o}dinger Equation within the
uniform flat space within sufficiently high potential barriers on both sides
of the surface, and then squeeze the width of barriers. Since the differnce
between the excited and the ground state energy of the particle along the
direction normal to the surface is very much larger than that of the
particle along the tangential direction so that the degree of freedom along
the normal direction is actually frozen to the ground state, an effective
dynamics for the constrained system on the surface is thus established. This
CPF has a distinct feature for no presence of any ambiguity. It is thus a
powerful tool to examine various curvature-induced consequences in
two-dimensional curved surfaces or curved wires \cite{packet0}. Experimental
confirmations include: an optical realization of the geometric potential 
\cite{exp1} in 2010 and the geometric potential in a one-dimensional
metallic $\mathit{C}_{60}$ polymer with an uneven periodic peanut-shaped
structure in 2012 \cite{exp2}. Applying the CPF to momentum operators which
are fundamentally defined as generators of a space translation, we have
geometric momenta \cite{liu13-2} which depends on the extrinsic curvatures
of the curved surface.

It is generally accepted that the canonical quantization offers a
fundamental framework to directly construct the quantum operators, where the 
\textit{fundamental quantum conditions} refer to a set of commutators\
between components of position and momentum \cite{dirac1,dirac2}. Many
explorations have been devoted to searching for the \textit{geometric
potential }within the framework \cite%
{geoquan,homma,ogawa,ikegami,Jaffe,mbbj,ILNC,kleinert,Tagirov,hong,Golovnev,2011,weinberg}%
. It is curious that no attempt is successful for even simplest
two-dimensional curved surface $\Sigma ^{2}$ embedded in $R^{3}$. Some
results are contradictory with each other \cite%
{homma,kleinert,hong,ogawa,ikegami,liu13-1}. We revisited all these
attempts, and concluded that the canonical quantization together with Schr%
\"{o}dinger-Podolsky-DeWitt approach of Hamiltonian operator construction
was dubious, for the kinetic energy in it takes some presumed forms of
distributing positions and momenta in the Hamiltonian. Since 2011, we have
tried to enlarge the canonical quantization scheme to simultaneously
quantize the Hamiltonian together with positions and momenta \cite%
{liu13-2,liu13-1,ECQ}, rather than substitute the position and momentum
operators into the presumed forms of Hamiltonian. Yet the success is limited
because of the operator-ordering problem \cite{liu13-2,liu13-1,ECQ,liu17}.
It seems that the operator-ordering problem is inherent to the quantization
and is hardly avoidable \cite{tdlee,dirac3}. For instance, apart from the
problem occurs in Hamiltonian, it even appears in the \textit{fundamental
quantum conditions, }see the last expression in (\ref{pp}). The key finding
of the present study is to identify existence of the operator-ordering-free
section in the enlarged scheme of quantum conditions, sufficiently to lead
to the geometric potential.

\section{Dirac brackets and quantization conditions}

Let us consider a non-relativistically free particle that is constrained to
remain on a surface $\Sigma ^{N-1}$ described by a constraint in the
configurational space $f(x)=0$, where $f(x)$ is some smooth function of
position $x$ in $R^{N}$, whose normal vector is $\mathbf{n}\equiv \nabla
f(x)/|\nabla f(x)|$. We can always choose the equation of the surface such
that $|\nabla f(x)|=1$ \cite{ikegami}, so that $\mathbf{n}\equiv \nabla f(x)$%
. This is because no matter what form of the surface equation we begin with,
only the unit normal vector and/or its derivatives enter the physics
equation. In classical mechanics, the Hamiltonian is simply $H=p^{2}/2\mu $
where $p$ denotes the momentum, and $\mu $ denotes the mass. However, in
quantum mechanics, we can not impose the usual canonical commutation
relations $[x_{i},p_{j}]=i\hbar \delta _{ij}$, ($i,j=1,2,3,...N$). Dirac
devised a general prescription to eliminate the motion in the direction
normal to the surface by introducing the Dirac brackets \cite{dirac2}, $%
[f(x,p),g(x,p)]_{D}\equiv \lbrack f(x,p),g(x,p)]_{P}-[f(x,p),\chi _{\alpha
}(x,p)]_{P}C_{\alpha \beta }^{-1}[\chi _{\beta }(x,p),g(x,p)]_{P}$, where $%
[f(x,p),g(x,p)]_{P}$ stands for the Poisson brackets, and repeated indices
are summed over in whole of this Letter, and $C_{\alpha \beta }\equiv
\lbrack \chi _{\alpha }(x,p),\chi _{\beta }(x,p)]_{P}$ are the matrix
elements in the constraint matrix $\left( C_{\alpha \beta }\right) $ and the
functions $\chi _{\alpha }(q,p)$ ($\alpha ,\beta =1,2$) are two constraints 
\cite{weinberg}, 
\begin{equation}
\chi _{1}(x,p)\equiv f(x)\left( =0\right) ,\text{and }\chi _{2}(x,p)\equiv 
\mathbf{n\cdot p}\left( =0\right) \mathbf{.}  \label{constraints}
\end{equation}%
It is an easy task to give following elementary Dirac brackets \cite%
{weinberg}, 
\begin{equation}
\lbrack x_{i},x_{j}]_{D}=0,[x_{i},p_{j}]_{D}=\delta
_{ij}-n_{i}n_{j},[p_{i},p_{j}]_{D}=(n_{j}n_{i,k}-n_{i}n_{j,k})p_{k},
\label{pp}
\end{equation}%
and \cite{weinberg,liu17}, 
\begin{equation}
\frac{d\mathbf{x}}{dt}\equiv \lbrack \mathbf{x},H]_{D}=\frac{\mathbf{p}}{\mu 
},\frac{d\mathbf{p}}{dt}\equiv \lbrack \mathbf{p},H]_{D}=-\frac{\mathbf{n}}{%
\mu }(\mathbf{p\cdot \nabla n\cdot p}).  \label{ph}
\end{equation}%
The Dirac bracket quantization hypothesizes that the definition of a quantum
commutator for any pair of variables $f$ and $g$ is given by \cite%
{dirac1,dirac2,dirac3}, 
\begin{equation}
\lbrack f,g]=i\hbar O\left\{ [f,g]_{D}\right\}  \label{quantization}
\end{equation}%
in which $O\left\{ f\right\} $ stands for the quantum operator corresponding
to the classical quantity $f$. The \textit{fundamental quantum conditions}
are widely taken to be comprised by following commutators $%
[x_{i},x_{j}],[x_{i},p_{j}]$ and $[p_{i},p_{j}]$. It must be mentioned that
Dirac himself had never assumed so except when the particle moves in flat
space where we should directly quantize the Poisson brackets, and he was
clearly aware of the operator-ordering difficulties which should be
carefully got over \cite{dirac3}. If taking the straightforward definition
of quantum condition for different components of the momentum, $%
[p_{i},p_{j}] $, we encounter a disturbing operator-ordering problem in $%
O\left\{ (n_{j}n_{i,k}-n_{i}n_{j,k})p_{k}\right\} $ \cite%
{homma,ogawa,ikegami}. Much more annoying operator-ordering problem appears
in $O\left\{ \mathbf{n}(\mathbf{p\cdot \nabla n\cdot p})\right\} $ if one
attempts to construct a quantum condition for $[\mathbf{p},H]$ \cite{liu17}.

However, in classical mechanics, we have following trivial consequences of (%
\ref{pp}) and (\ref{ph}), respectively, 
\begin{equation}
\mathbf{n}\cdot \lbrack \mathbf{x},H]_{D}=\mathbf{n}\cdot \frac{\mathbf{p}}{%
\mu }=0,\text{ and }\mathbf{n}\wedge \lbrack \mathbf{p},H]_{D}=0.
\label{zero}
\end{equation}%
The first equation shows that the motion lies in the tangential plane, and
the second shows that particle experiences no tangential force. The first
equation is a single one, and the second one $\mathbf{n}\wedge \lbrack 
\mathbf{p},H]_{D}=0$ has $N(N-1)/2$ independent equations for its component
form is $\varepsilon _{\lbrack ij]}n_{i}[p_{j},H]=0,(i\neq j)$ where $%
\varepsilon _{\lbrack ij]}\equiv \varepsilon _{i_{1}i_{2}...i...j...i_{N}}$
that is the Levi-Civita symbol of rank $N$, and the positions of two indexes 
$ij$ ($j\succ i$) in the array $i_{1}i_{2}...i...j...i_{N}$ of $\varepsilon
_{\lbrack ij]}$ are arbitrary.

In order to transit to quantum mechanics for the system under study, the
section of quantum conditions\textit{\ free from the operator-ordering
difficulty} is given by,%
\begin{eqnarray}
&&[x_{i},x_{j}]=0,\text{ }[x_{i},p_{j}]=i\hbar \left( \delta
_{ij}-n_{i}n_{j}\right) ,\text{ }[\mathbf{x},H]=i\hbar \frac{\mathbf{p}}{\mu 
},  \label{zero1} \\
&&\varepsilon _{\lbrack ij]}\left( n_{i}[p_{j},H]+[p_{j},H]n_{i}\right)
=0,(i\neq j).  \label{zero2}
\end{eqnarray}%
The first finding of this Letter is that this set (\ref{zero1})-(\ref{zero2}%
) is identified as the operator-ordering-free section of the enlarged set of
quantum conditions defined by all commutators between $\{\mathbf{x},\mathbf{p%
},H\}$. To clearly demonstrate the critical importance of the "trivial"
relation (\ref{zero2}), let us assume that the Hamiltonian operator is given
by,%
\begin{equation}
H=-\frac{\hbar ^{2}}{2\mu }\nabla _{LB}^{2}+V_{G},  \label{trialh}
\end{equation}%
where $V_{G}$ is the curvature-induced potential, and $\nabla _{LB}^{2}={%
\nabla _{S}}\cdot {\nabla _{S}}$ is the Laplace-Beltrami operator which is
the dot product of the gradient operator ${\nabla _{S}\equiv }\mathbf{e}%
_{i}(\delta _{ij}-n_{i}n_{j})\partial _{j}$ $=\nabla _{N}-\mathbf{n}\partial
_{n}$ on the surface $\Sigma ^{N-1}$ with $\nabla _{N}$ being usual gradient
operator in $R^{N}$ \cite{ikegami}. The relation between the
Laplace-Beltrami operator $\nabla _{LB}^{2}$ on $\Sigma ^{N-1}$ and the
usual Laplacian operator $\Delta _{N}\equiv \partial _{i}\partial _{i}$ in $%
R^{N}$ is $\nabla _{LB}^{2}={\nabla _{S}\cdot \nabla _{S}}$ $=\partial
_{i}(\delta _{ij}-n_{i}n_{j})\partial _{j}=\Delta _{N}+M\partial
_{n}-\partial _{n}^{2}$ with ${M}$ denoting the mean curvature that is in
fact the trace of the extrinsic curvature tensor \cite{ikegami}. It is
straightforward to show that relations $\mathbf{p=}$ $[\mathbf{x},H](\mu
/i\hbar )$ in (\ref{zero1}) give the geometric momentum \cite{liu13-1},%
\begin{equation}
{\mathbf{p}}=-i\hbar ({\nabla _{S}}+\frac{{M{\mathbf{n}}}}{2}).  \label{GM}
\end{equation}%
This geometric momentum is in fact the hermitian operator corresponding to $%
-i\hbar {\nabla _{S}\equiv }-i\hbar (\nabla _{N}-\mathbf{n}\partial _{n})$.
In consequence, the commutators $[p_{i},p_{j}]$ turn out to satisfy the
following relation with $f,_{k}\equiv \partial f/\partial x_{k}$, 
\begin{equation}
\lbrack p_{i},p_{j}]=\frac{i\hbar }{2}\left(
(n_{j}n_{i,l}-n_{i}n_{j,l})p_{l}+p_{l}(n_{j}n_{i,l}-n_{i}n_{j,l})\right) ,
\label{properpp}
\end{equation}%
and the Hamiltonian operator turns out to be,%
\begin{equation}
H=\frac{p^{2}}{2\mu }-\frac{\hbar ^{2}}{8\mu }{M}^{2}+V_{G}.  \label{trialh2}
\end{equation}%
The second and key finding of this Letter is: With the quantum condition (%
\ref{zero2}) being imposed, $V_{G}$ must be the expected geometric potential 
\cite{dacosta,2004,packet0,ogawa,ikegami,Jaffe} with $K\equiv \left(
n_{i,j}\right) ^{2}$ that is in fact the trace of square of the extrinsic
curvature tensor \cite{ikegami}, 
\begin{equation}
V_{G}=-\frac{\hbar ^{2}}{4\mu }K+\frac{\hbar ^{2}}{8\mu }{M}^{2}.  \label{GP}
\end{equation}

\textit{Proof: }To note that commutator $[p_{j},H]$ is,%
\begin{equation}
\lbrack p_{j},H]=\frac{1}{2\mu }[p_{j},p_{k}p_{k}]+[p_{j},W]=\frac{1}{2\mu }%
\left( [p_{j},p_{k}]p_{k}+p_{k}[p_{j},p_{k}]\right) +[p_{j},W],
\end{equation}%
where $W\equiv V_{G}-{M}^{2}\hbar ^{2}/(8\mu )$. Now we examine the
expression $[p_{j},p_{k}]p_{k}+p_{k}[p_{j},p_{k}]\equiv F_{j}+G_{j}$, and
substituting relation (\ref{properpp}) into it, we have for $F_{j}$ and $%
G_{j}$ respectively,%
\begin{eqnarray}
F_{j} &\equiv &\frac{i\hbar }{2}\left\{
n_{j,l}n_{k}p_{l}p_{k}+p_{l}n_{j,l}n_{k}p_{k}+p_{k}n_{k}n_{j,l}p_{l}+p_{k}p_{l}n_{j,l}n_{k}\right\}
\label{Fj} \\
&\overset{c.l.}{\Rightarrow }&\frac{i\hbar }{2}\left\{ {\mathbf{p}}\mathbf{%
\cdot }\nabla n_{j}\mathbf{n\cdot }{\mathbf{p}}+{\mathbf{p}}\mathbf{\cdot }%
\nabla n_{j}\mathbf{n\cdot }{\mathbf{p}}+{\mathbf{p}}\mathbf{\cdot }\nabla
n_{j}\mathbf{n\cdot }{\mathbf{p}}+{\mathbf{p}}\mathbf{\cdot }\nabla n_{j}%
\mathbf{n\cdot }{\mathbf{p}}\right\} =2i\hbar {\mathbf{p}}\mathbf{\cdot }%
\nabla n_{j}\left( \mathbf{n\cdot }{\mathbf{p}}\right) , \\
G_{j} &\equiv &-\frac{i\hbar }{2}\left\{
n_{j}n_{k,l}p_{l}p_{k}+p_{l}n_{j}n_{k,l}p_{k}+p_{k}n_{j}n_{k,l}p_{l}+p_{k}p_{l}n_{j}n_{k,l}\right\}
\label{Gj} \\
&\overset{c.l.}{\Rightarrow }&-\frac{i\hbar }{2}\left\{ n_{j}{\mathbf{p}}%
\mathbf{\cdot }\nabla \mathbf{n\cdot }{\mathbf{p}}+n_{j}{\mathbf{p}}\mathbf{%
\cdot }\nabla \mathbf{n\cdot }{\mathbf{p}}+n_{j}{\mathbf{p}}\mathbf{\cdot }%
\nabla \mathbf{n\cdot }{\mathbf{p}}+n_{j}{\mathbf{p}}\mathbf{\cdot }\nabla 
\mathbf{n\cdot }{\mathbf{p}}\right\} =-2i\hbar n_{j}{\mathbf{p}}\mathbf{%
\cdot }\nabla \mathbf{n\cdot }{\mathbf{p,}}
\end{eqnarray}%
where $c.l.$ denotes the classical limit. Clearly, $F_{j}$ (\ref{Fj})
vanishes in classical mechanics for $\mathbf{n\cdot }{\mathbf{p}}=0$, while $%
G_{j}$ (\ref{Gj}) corresponds to the centripetal force $-2n_{j}{\mathbf{p}}%
\mathbf{\cdot }\nabla \mathbf{n\cdot }{\mathbf{p}}$. Thus, in the classical
limit, the commutators $[\mathbf{p},H]$ is $-i\hbar \mathbf{n}\left( \mathbf{%
p\cdot \nabla n\cdot p}\right) /\mu $, but in general it never be $-i\hbar
O\left\{ \mathbf{n}\left( \mathbf{p\cdot \nabla n\cdot p}\right) \right\}
/\mu $ because $[\mathbf{p},H]$ contains both the geometric potential that
proportional to $\hbar ^{2}$ and terms proportional to $\mathbf{n\cdot }{%
\mathbf{p}}$ that vanishes only in classical limit. In left-handed side of
Eq. (\ref{zero2}), we need to deal with $\varepsilon _{\lbrack ij]}\left(
n_{i}F_{j}+F_{j}n_{i}\right) $ and $\varepsilon _{\lbrack ij]}\left(
n_{i}G_{j}+G_{j}n_{i}\right) $, respectively. After somewhat lengthy but
straightforward calculations, we find heavy cancellations among terms, and
we find a very simple result,%
\begin{equation}
\varepsilon _{\lbrack ij]}\left( n_{i}F_{j}+F_{j}n_{i}\right) =\varepsilon
_{\lbrack ij]}\left( n_{i}G_{j}+G_{j}n_{i}\right) =-\left( i/2\right) \hbar
\varepsilon _{\lbrack ij]}\left( \hbar ^{2}n_{i}K,_{j}\right) .
\end{equation}%
Next, we compute $\varepsilon _{\lbrack ij]}\left(
n_{i}[p_{j},W]+[p_{j},W]n_{i}\right) $ in Eq. (\ref{zero2}), which can be
shown to be,%
\begin{equation}
\varepsilon _{\lbrack ij]}\left( n_{i}[p_{j},W]+[p_{j},W]n_{i}\right)
=-2i\hbar \varepsilon _{\lbrack ij]}n_{i}W,_{j}.
\end{equation}%
The Eq. (\ref{zero2}) is then, 
\begin{equation}
\varepsilon _{\lbrack ij]}\left( n_{i}[p_{j},H]+[p_{j},H]n_{i}\right)
=-2i\hbar \varepsilon _{\lbrack ij]}n_{i}\left( \frac{\hbar ^{2}}{4\mu }%
K+W\right) ,_{j}=0.
\end{equation}%
It is an orthogonal relation, which is in vector form, 
\begin{equation}
-2i\hbar \mathbf{n}\wedge \mathbf{h}=0,
\end{equation}%
where $\mathbf{h}$ must be in parallel with normal $\mathbf{n}$ itself, and
for convenience we assume $\mathbf{h}=\varphi (x)$ $\nabla f(x)$ where $%
\varphi (x)$ is an arbitrary function that can never be zero on any point of
the surface $f(x)=0$ that immediately leads to relation $\nabla \left(
\varphi (x)f(x)\right) =\varphi (x)\nabla f(x)$. In final, we obtain,%
\begin{equation}
\frac{\hbar ^{2}}{4\mu }K+W=\varphi (x)f(x)+const.=const.\text{ }I.e.\text{ }%
W=const.-\frac{\hbar ^{2}}{4\mu }K.  \label{final}
\end{equation}%
Recalling $W\equiv V_{G}-{M}^{2}\hbar ^{2}/(8\mu )$, we find that the
geometric potential $V_{G}$ given by (\ref{final}) differs from (\ref{GP})
by a constant which can be set to be zero. \textit{Q.E.D.}

\section{Conclusions and discussions}

The quantum conditions given by the straightforward applications of the
quantization rule (\ref{quantization}) are not always fruitful, even
misleading. For the particle on the curved surface, in order to obtain the
geometric potential predicted by the so-called CPF, a proper enlargement of
the quantum conditions turns out to be compulsory to contain positions,
momenta, and Hamiltonian. What is more, a construction of unambiguous
quantum conditions out of the equation (\ref{quantization}) proves
inevitable. Combining the\ enlargement and the construction, we successfully
obtain the geometric potential. Thus, the fundamental problem regarding the
Dirac quantization of a free particle on a curved hypersurface is now
resolved.

Finally, we would like to point out two points: 1, We do not know yet
whether the operator-ordering-free section of the enlarged set of quantum
conditions always exists in general, and if finding one, we will fix the
Lee's operator-ordering problem \cite{tdlee}. 2, There are other forms of
the enlargement and the construction of the quantum conditions in
literature, for instance Refs. \cite{bender,Deriglazov}, but they took
complete different forms and were devised to serve entirely different
purposes.

\begin{acknowledgments}
This work is financially supported by National Natural Science Foundation of
China under Grant No. 11675051.
\end{acknowledgments}


\begin{thebibliography}{99}
\bibitem{dewitt} B. S. DeWitt, Rev. Mod. Phys. \textbf{29}, 377(1957).{}

\bibitem{Sch} E. Schr\"{o}dinger, Ann. Phys. (Leipzig), \textbf{79},
734(1926).

\bibitem{podolsky} B. Podolsky, Phys. Rev. \textbf{32}, 812(1928).

\bibitem{jk} H. Jensen and H. Koppe, Ann. Phys. \textbf{63}, 586(1971).

\bibitem{dacosta} R. C. T. da Costa, Phys. Rev. A \textbf{23}, 1982(1981), 
\textbf{25} 2893(1982).

\bibitem{2004} A. V. Chaplik and R. H. Blick, New J. Phys. \textbf{6},
33(2004).

\bibitem{packet0} M. V. Entin, and L. I. Magarill, Phys. Rev. B \textbf{64},
085330(2001); G. Ferrari and G. Cuoghi, Phys. Rev. Lett. \textbf{100},
230403 (2008); V. Atanasov, R. Dandoloff, and A. Saxena, Phys. Rev. B 
\textbf{79}, 033404(2009); Y. N. Joglekar and A. Saxena, Phys. Rev. B 
\textbf{80}, 153405(2009); S. Ono and H. Shima, Phys. Rev. B \textbf{79},
235407(2009); H. Shima, H. Yoshioka, and J. Onoe, Phys. Rev. B \textbf{79,}
201401(2009); V. Atanasov and A. Saxena, Phys. Rev. B \textbf{81,}
205409(2010); S. Ono, H. Shima, Physica E: Low Dimens. Syst. Nanostruct. 
\textbf{42}, 1224(2010); F. T. Brandt, J. A. S\'{a}chez-Monroy, Europhys.
Lett. \textbf{111}, 67004(2015); C. Ortix and J. van den Brink, Phys. Rev. B 
\textbf{83}, 113406(2011); D. Schmeltzer, J. Phys. Condens. Matter \textbf{23%
}, 155601(2011); E. O. Silva, S. C. Ulhoa, F. M. Andrade, C. Filgueiras, R.
G. G. Amorim, Ann. Phys. (N.Y.), \textbf{362}, 739(2015); C. Filgueiras, E.
O. Silva, Phys. Lett. A \textbf{379},\textbf{\ }2110(2015); Y. L. Wang, L.
Du, C. T. Xu, X. J. Liu, H. S. Zong, Phys. Rev. A, \textbf{90} 042117(2014);
Y. L. Wang, H. S. Zong, Ann. Phys. (N.Y.), \textbf{364} 68(2016); L. Du, Y.
L. Wang, G. H. Liang, G. Z. Kang, X. J. Liu, H. S. Zong, Physica E: Low
Dimens. Syst. Nanostruct. \textbf{76,} 28(2016); Y. L. Wang, H. Jiang, and
H. S. Zong, arXiv:quant-ph/1702.00893v1 (2017).

\bibitem{exp1} A. Szameit, F. Dreisow, M. Heinrich, R. Keil, S. Nolte, A. T%
\"{u}nermann, and S. Longhi, Phys. Rev. Lett. \textbf{104},\textbf{\ }%
150403(2010).

\bibitem{exp2} J. Onoe, T. Ito, H. Shima, H. Yoshioka, and S. Kimura,
Europhys. Lett. \textbf{98, }27001(2012).

\bibitem{liu13-2} Q. H. Liu, L. H. Tang, D. M. Xun, Phys. Rev. A \textbf{84}%
, 042101(2011), Q. H. Liu, J. Phys. Soc. Jpn. \textbf{82}, 104002(2013).

\bibitem{dirac1} P. A. M. Dirac, \textit{The Principles of Quantum Mechanics}%
, 4th ed. (Oxford University Press, Oxford, 1967)P.114.

\bibitem{dirac2} P. A. M. Dirac, \textit{Lectures on quantum mechanics}
(Yeshiva University, New York, 1964); Can. J. Math. \textbf{2}, 129(1950).

\bibitem{geoquan} J. \'{S}niatycki, Geometric \textit{Quantization and
Quantum Mechanics} (Springer--Verlag, New York, Heidelberg, Berlin, 1980).

\bibitem{homma} T. Homma, T. Inamoto, T. Miyazaki, \textit{Phys. Rev. D} 
\textbf{42}, 2049(1990).

\bibitem{ogawa} N. Ogawa, K Fujii and A. Kobushkin, Prog. Theor. Phys. 
\textbf{83}, 894(1990). N. Ogawa, K. Fujii, N. Chepilko and A. Kobushkin,
Prog. Theor. Phys. \textbf{85}, 1189(1991), N. Ogawa, Progress Theor. Phys. 
\textbf{87}, 513(1992).

\bibitem{ikegami} M. Ikegami, Y. Nagaoka, S. Takagi, and T. Tanzawa,\textit{%
\ }Prog. Theor. Phys. \textbf{88}, 229(1992).

\bibitem{Jaffe} P. C. Schuster, and R. L. Jaffe, Ann. Phys. \textbf{307},%
\textbf{\ }132-143(2003).

\bibitem{mbbj} M. Burgess and B. Jensen, Phys. Rev. A, \textbf{48,}
1861(1993).

\bibitem{ILNC} C. Destri, P. Maraner, E. Onofri, Nuov. Cim. A, \textbf{107},
237(1994).

\bibitem{kleinert} H. Kleinert and S. V. Shabanov, Phys. Lett. A \textbf{232}%
, 327(1997).

\bibitem{Tagirov} E. A. Tagirov, arXiv:quant-ph/0101016v1 (2001); N. Ogawa,
arXiv:hep-th/9703181v3 (1997).

\bibitem{hong} S. T. Hong, and K. D. Rothe, Ann. Phys. \textbf{311},
417(2004).

\bibitem{Golovnev} A. V. Golovnev, Rep. Math. Phys. \textbf{64}, 59(2009).

\bibitem{2011} B. Jensen, R. Dandoloff, Phys. Lett. A \textbf{375},
448(2011).

\bibitem{weinberg} S. Weinberg, \textit{Lectures on Quantum Mechanics}, 2nd
ed., (Cambridge University Press, Cambridge, 2015).

\bibitem{liu13-1} Q.H. Liu, J. Math. Phys. \textbf{54}, 122113(2013).

\bibitem{ECQ} D. M. Xun, Ann. Phys. (N.Y.) \textbf{338,} 123(2013); D. M.
Xun and Q. H. Liu, Ann. Phys. (N.Y.) \textbf{341,} 132(2014); D. M. Xun, and
Q. H. Liu, Int. J. Geom. Meth. Mod. Phys. \textbf{10,} 1220031(2013); Z. S.
Zhang, S. F. Xiao, D. M. Xun and Q. H. Liu, Commun. Theor. Phys. \textbf{63,}
19(2015).

\bibitem{liu17} Q. H. Liu, J. Zhang, D.K. Lian, L. D. Hu and Z. Li, Physica
E: Low Dimens. Syst. Nanostruct. \textbf{87}, 123(2017), L. D. Hu, D. K.
Lian, Q. H. Liu, Eur. Phys. J. C \textbf{76,} 655(2016).

\bibitem{tdlee} T. D. Lee, \textit{Particle Physics and Introduction to
Field Theory, }Vol.1, (Harwood Academic Publishers GmbG, N.Y., 1981) p.5.

\bibitem{dirac3} P. A. M. Dirac, Proc. Royal Soc. (London) A, \textbf{109},
642(1925).

\bibitem{bender} C. M. Bender, and G. V. Dunne, Phys. Rev. D \textbf{40},
2739(1989).

\bibitem{Deriglazov} A. Deriglazov, A. Nersessian, Phys. Lett A \textbf{378}%
, 1224(2014).
\end{thebibliography}
\end{document}